\documentclass[aps,pra,epsfigure,twocolumn,longbibliography,superscriptaddress]{revtex4-1}

\usepackage{dcolumn}    
\usepackage{bm} 
\usepackage{graphicx}
\usepackage{amsmath}    
\usepackage{latexsym}
\usepackage{amsfonts}   
\usepackage{amssymb}
\usepackage{array}      
\usepackage{epsfig}
\usepackage{txfonts}
\usepackage{xcolor}
\usepackage[colorlinks=true,linkcolor=blue,urlcolor=blue,citecolor=blue,pdfusetitle]{hyperref}
\usepackage{hyperref}
\usepackage{ulem}

\newcommand{\ket}[1]{{\left\vert {#1} \right\rangle}}

\newcommand{\ketbra}[2]{|{#1}\rangle\langle{#2}|}

\newcommand{\beq}{\begin{equation}}
\newcommand{\eeq}{\end{equation}}
\newcommand{\bea}{\begin{eqnarray}}
\newcommand{\eea}{\end{eqnarray}}
\begin{document}
\author{Eoghan Ryan}
\affiliation{Centre for Theoretical Atomic, Molecular, and Optical Physics, School of Mathematics and Physics, Queen's University, Belfast BT7 1NN, United Kingdom}
\author{Mauro Paternostro}
\affiliation{Centre for Theoretical Atomic, Molecular, and Optical Physics, School of Mathematics and Physics, Queen's University, Belfast BT7 1NN, United Kingdom}
\author{Steve Campbell}
\affiliation{School of Physics, University College Dublin, Belfield, Dublin 4, Ireland}
\affiliation{Centre for Quantum Engineering, Science, and Technology, University College Dublin, Belfield, Dublin 4, Ireland}

\title{%Mauro's Onion:\\
Quantum Darwinism in a structured spin environment}

\begin{abstract}
We examine the emergence and suppression of signatures of quantum Darwinism when the system of interest interacts with a complex, structured environment. We introduce an extended spin-star model where the system is coupled to $N$ independent spin-chains. Each site of the chain then lives in a definite layer of the environment, and hence we term this the ``onion" model. We fix the system-environment interaction such that classical objectivity is guaranteed if the environment consists of a single layer. Considering a fully factorized initial state for all constituent sub-systems, we then examine how the emergence and proliferation of signatures of quantum Darwinism are delicately dependent on the chain interaction, establishing that when the chains are considered as indivisible fragments to be interrogated, characteristic redundancy plateaux are always observed at least transiently. In contrast, observing a redundant encoding in a specific layer is highly sensitive to the nature of the interaction. Finally, we consider the case in which each chain is initialized in the ground state of the interaction Hamiltonian, establishing that this case shares the qualitative features of the factorized initial state case, however now the strength of the applied magnetic field has a significant impact on whether quantum Darwinism can be observed. We demonstrate that the presence or absence of quantum Darwinistic features can be understood by analysing the correlations within a layer using total mutual information and global quantum discord.
\end{abstract}
\maketitle

\section{Introduction}
The concept of decoherence notoriously captures how inherently quantum properties, from quantum coherence to entanglement and non-locality, are lost through interactions with an environment~\cite{ZurekRMP}. This mechanism leads to the establishment of crucial system-environment correlations, which are responsible for the degradation of the information initially encoded in the state of the system. However, the framework of decoherence alone does not capture the microscopic features of the quantum-to-classical transition from which classical objectivity emerges. 

Quantum Darwinism~\cite{ZurekRMP, ZurekNatPhys2009, ZwolakPRA2010, ZwolakPRL2009, ZurekPRA2017, ZurekSciRep2016, GiorgiPRA2015, GarrawayPRA2017, KohoutPRA2006, NadiaPRA, DarwinismExp1, DarwinismExp2, DarwinismExp3, SalvatorePRR, SabrinaNPJQI, SabrinaPRR, BalaneskovicaEPJD2015, MendlerEPJD2016, KnottQD, CampbellEntropy, ZwolakNJP2012, AdessoPRL2018} attempts at providing this by elevating the role of the environment from a monolithic sink into a fragmented information storage space. Different observers can then independently access their part of the environment, and perform measurements, whose outcomes they compare. In the Darwinistic picture the objective, classical description of the system emerges from quantum mechanics owing to the proliferation of redundant information throughout the environment resulting from the system-environment correlations established by their mutual interaction. Key to quantum Darwinism is that the states produced by environment-induced decoherence encode local copies of classical information about the system, whose proliferation allows  the observers holding different parts of the environment to agree on the quantity of information they have on the system~\cite{ZurekRMP, ZurekNatPhys2009}. This occurs when measurements are performed in the pointer basis~\cite{ZurekRMP}, i.e.  the basis of an operator that commutes with the one  governing the system-environment interactions. This basis is the one in which mutual information is generally maximised and results from the process of  einselection~\cite{ZurekRMP}. Thus, it is important to stress that the framework of quantum Darwinism should not be confused with classical and/or semi-classical limits corresponding to large system sizes. Rather, quantum Darwinism deals with the mechanism by which information about a system state is encoded into its environment, and therefore one can readily examine the pertinent features provided the environment consists of as little as three constituents. Indeed recently, tests of quantum Darwinism have been carried in a variety of set-ups. In particular, photonic cluster states~\cite{DarwinismExp1}, a photon simulator~\cite{DarwinismExp2} and nitrogen-vacancy centers~\cite{DarwinismExp3}, all of which experimentally verify the predictions made by quantum Darwinism. 

The key figure of merit in the characterization of quantum Darwinism is the quantum mutual information between the system of interest $S$ and a fragment of  environment ${\cal E}_f$~\cite{ZurekRMP, ZurekNatPhys2009}
\begin{equation}
\label{MutInfo}
I(S:\mathcal{E}_{f})=H(\rho_S) +H(\rho_{\mathcal{E}_{f}})-H(\rho_S,\rho_{\mathcal{E}_{f}})
\end{equation}
where $H(\cdot)$ denotes the von Neumann entropy and $\rho_S$ ($\rho_{\mathcal{E}_{f}}$) stands for the density matrix of the system (fragment of environment with $f\in [0,1]$ quantifying the fraction of the environment this fragment represents). When $\mathcal{I}(S:\mathcal{E}_{f})=H(S)$, the information of the system is stored completely in $\mathcal{E}_{f}$. Given a large enough number of fragments where this is the case, then the information of the system is said to have redundantly proliferated into the surrounding environment and a redundancy plateau emerges in the mutual information, where the further consideration of more fragments of the environment, or increase in the size of the fragment under consideration, will not reveal any further information regarding the system of interest. This plateau will continue until the fragment of the environment encompasses the whole environment and the mutual information rises to $2H(S)$. As such, it is often more convenient to work with a rescaled mutual information
\beq
\label{MIeq}
\overline{\mathcal{I}} = \frac{\mathcal{I}(S:\mathcal{E}_f)}{H(S)}
\eeq

In this work we examine how signatures of classical objectivity are affected when the environment is a complex entity. In particular we introduce an extended spin-star model~\cite{SpinStar, CampbellPRA2019}, termed the ``onion model", where the system of interest interacts with only the inner most layer of the environment. The constituents of this inner layer then couple to individual qubits in the next layer via some suitable interaction term and so on. Thus, the system is in effect coupled to several independent spin chains. We establish that, while for an environment that consists of a single layer, the redundant encoding can be readily observed for suitable interactions~\cite{CampbellPRA2019}, signatures of objectivity do not easily proliferate in a more complex environment. In particular, we show that the nature of the intra-layer interactions leads to highly non-trivial dynamics of the mutual information shared between the system and the environmental fragments, with redundancy plateaux emerging under certain specialised conditions. We establish that if entire chains are treated as indivisible environment fragments, that quantum Darwinism will faithfully emerge. However, the nature of the intra-layer interactions can significantly affect the temporal windows in which the characteristic redundant encoding is observed. Finally we show that these behaviors can be understood by examining the dynamics of the quantum and total correlations shared within a given layer. 

Our work therefore complements previous studies which have explored the emergence of classical objectivity for different environmental configurations, such as a random unitary model~\cite{BalaneskovicaEPJD2015}, spin-$\tfrac{1}{2} XX$ models~\cite{GiorgiPRA2015}, bosonic environments~\cite{GarrawayPRA2017}, non-Markovianity~\cite{GalveSciRep2016, LewensteinPRA2017}, and collision models~\cite{SabrinaPRR, CampbellPRA2019}. In concordance with our results, these works all highlight that the microscopic details entering the precise physical description has a significant effect on whether or not signatures of quantum Darinwism are observed. However, our work provides additional insights beyond those reported in, for example Refs.~\cite{GiorgiPRA2015, BalaneskovicaEPJD2015, GalveSciRep2016, CampbellPRA2019, LewensteinPRA2017, SabrinaPRR}, with regards to the ability--or lack thereof--for the relevant system information to proliferate within the environment. While quantum Darwinism demonstrates neatly the means by which the system, through {\it direct} interaction with the environment, can redundantly encode information, we show that the nature of the intra-environment interactions can have a significant impact on whether and how this information is spread. In particular we show that an indirect redundant encoding of the system information, i.e. the establishment of suitable correlations between the system and environmental constituents that it has only indirectly communicated with, occurs in only special circumstances.

The remainder of this paper is organized as follows. In Sec.~\ref{onion} we present the model that we study and the quantitative tools that will be used in our analysis. Sec.~\ref{DarwinInTheOnion} is dedicated to the quantitative study of Darwinism in the layered structure at the core of our investigation. Finally, in Sec.~\ref{conc} we draw our conclusions and perspectives for further studies. 

\section{The Onion Model and correlation measures}
\label{onion}
\begin{figure}[t]
\includegraphics[width=0.95\columnwidth]{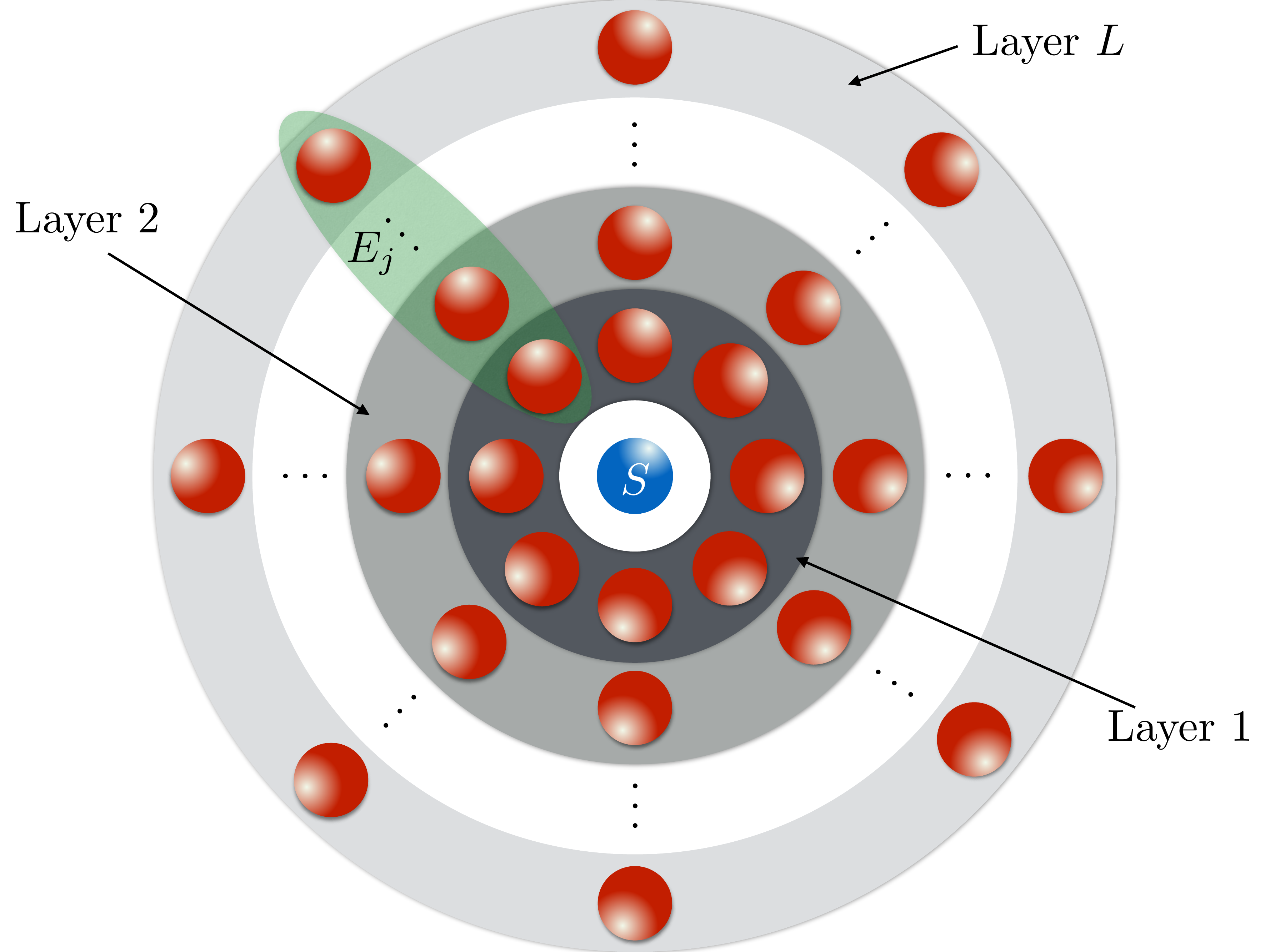}
\caption{Schematic of the ``onion" model to study the proliferation of classical objectivity. The environment is composed of several independent spin chains whose interactions are governed by a nearest neighbor anistropic Heisenberg interaction, Eq.~\eqref{subenvint}. The central system of interest interacts with the first subsystem of a given environmental chain via Eq.~\eqref{SysEnvInt}. We will assume that the system is initialised in $\ket{1}$ and we will consider two possibilities for the environment, either all qubits are identically prepared in $\ket{0}$ or each chain is the ground state of the interaction Hamiltonian.}
\label{schema}
\end{figure}

To explore the proliferation of system information into a complex environment we introduce the ``onion model" as shown in Fig.~\ref{schema} and, for simplicity, we assume that the system and all environmental units are qubits, however we expect our results to hold qualitatively for higher dimensional constituents. Here the central system of interest interacts with an environment that consists of $N$ independent spin chains $E_j$. The system only interacts with the first environmental layer. Thus, the model can be viewed as an extended spin-star model. Indeed, the emergence of classical objectivity for only a single layer (and therefore a standard spin-star system) was exhaustively explored in Ref.~\cite{CampbellPRA2019} where the nature of the interaction and initial conditions where shown to be crucial in establishing the emergence of quantum Darwinism. Therefore, in what follows we will assume that the interaction between the central system $S$ and the first qubit in a given environmental chain takes the form (here and throughout we assume units such that $\hbar\!=\!1$)
\begin{equation}
\label{SysEnvInt}
\mathcal{H}_{SE_j} = J \sigma^{S}_x \otimes \sigma^{j,1}_x.
\end{equation}
We assume that the system is always initially prepared in $\ket{1}$ -- with $\{\ket{0},\ket{1}\}$ the basis of the Pauli $z$ operator $\sigma_z$ -- and that the free Hamiltonian of the system $\mathcal{H}_S\!=\!\omega_S \sigma_z^S$ is such that $\omega_S\!\ll\!J$ and the effect of the free dynamics of the system can be neglected as compared to that of the interaction term with the environment. If the latter consists of only a single layer, with all qubits initialised in $\ket{0}$, this interaction leads to a dephasing of the system in the $\sigma_x$ basis and builds strong entanglement with the qubits in the environment which is necessary for a redundant encoding of the information~\cite{KorbiczPRA19} and the emergence of quantum Darwinism~\cite{CampbellPRA2019}. 

Our interest lies in whether, and how, a redundant encoding of the information can proliferate when the environment itself is a more complex entity. As previously indicated, in order to introduce more layers we assume that each qubit from the innermost layer constitutes the first qubit of a spin chain governed by the general anisotropic Heisenberg interaction
\begin{equation}
\label{subenvint}
\mathcal{H}_{E_j} = \sum_{k=1}^{L-1}\sum_{m=x,y,z}  \left(J_m \sigma^{j,k}_m \otimes \sigma^{j,k+1}_m \right) + B \sum_{k=1}^L \sigma^{j,k}_z.
\end{equation}
Here $\sigma_{m}^{j,k}$ are corresponding Pauli operators on the $k$-th qubit of sub-environment $E_j$, and $2B$ is the energy splitting between the energy levels of the particular environmental element. Crucially, we do not allow for qubits within a given layer to mutually interact, i.e. the chains $E_j$ do not directly interact. Therefore, the only way information can spread to outer layers is via the interaction within the sub-environments $E_j$. In what follows we explore how Eq.~\eqref{subenvint} dictates whether quantum Darwinism can be witnessed at the level of individual layers of the onion model and how the initialization of all environmental sub-units play a role.

We remark that the emergence of quantum Darwinism for structured environments has been explored previously~\cite{GiorgiPRA2015, GarrawayPRA2017}. An environment consisting of a single spin-ring governed by an $XX$ interaction was considered in Ref.~\cite{GiorgiPRA2015}, where it was shown that the strong correlations that the individual fragments of the environment can establish could hinder the emergence of objectivity. In contrast, our model explicitly forbids qubits within a given layer from directly coupling with one another. Ref.~\cite{GarrawayPRA2017} considered a similar model to the one considered presently which consisted of a system coupled to $N$ independent sub-environments, each comprising $j$ bosonic modes, with the system interacting only with the first mode. Emergence of objectivity was addressed when one considers the complete sub-environments as indivisible fragments and when one restricts to only the first mode. These situations being akin to considering the complete chains or only the first layer of our spin environment, respectively. 

While the nature of settings precludes us from drawing strong comparisons with such previous studies, in what follows, we nevertheless find a largely consistent picture. In particular, when the complete sub-environment is taken as a fragment, signatures of quantum Darwinism emerge, while restricting to special portions of the environment requires a much more careful analysis before one can argue genuine classical objectivity is witnessed. Additionally, our study extends the interesting results of Ref.~\cite{GarrawayPRA2017} by exploring the proliferation of objectivity within a structured environment. 

\begin{figure*}[t!]
\includegraphics[width=2.0\columnwidth]{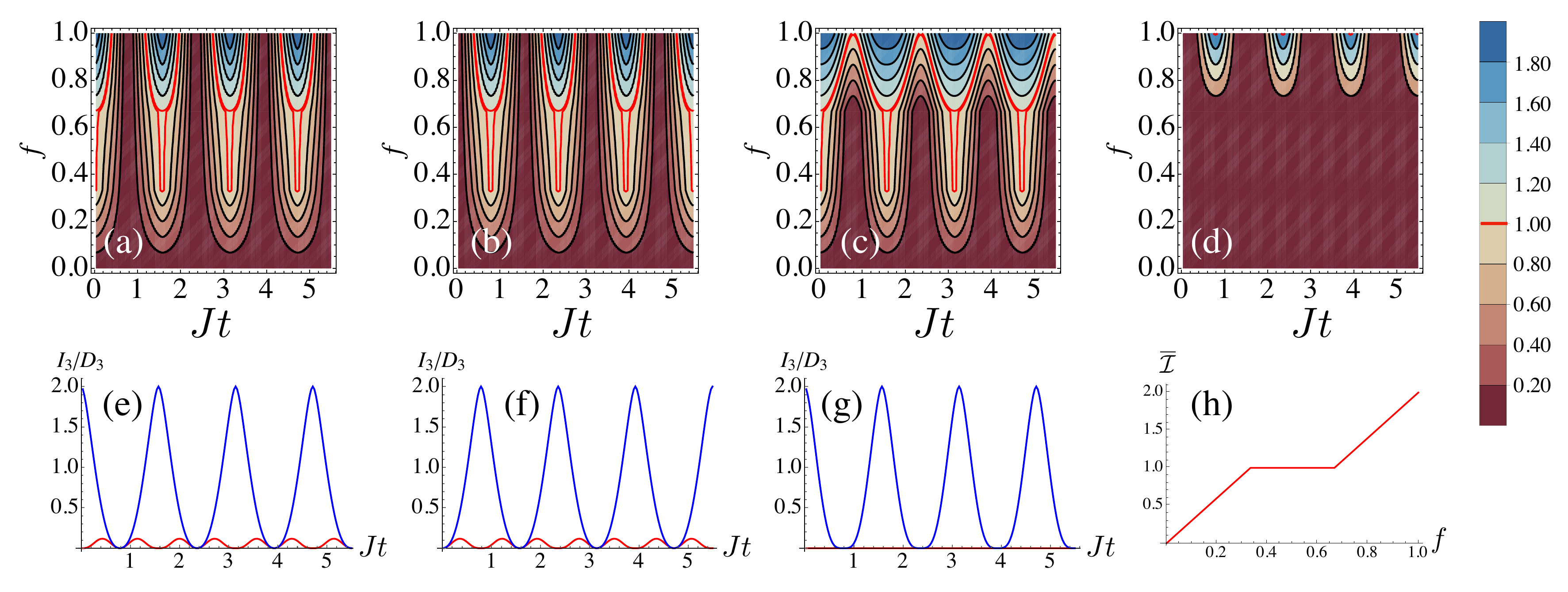}
\caption{Rescaled mutual information shared between the system and fragments of the environment consisting of (a,c) only the qubits in layer 1 and (b,d) only the qubits in layer 2, when the system-environment interaction occurs first and is subsequently switched off before the internal dynamics of the chains is switched on. (a,b) Corresponds to the case of an exchange interaction, $J_x\!=\!J_y=\!1$, $J_z\!=\!0$, and $B\!=\!0$, within the chains. (c,d) Corresponds to the case of an Ising interaction, $J_x\!=\!1$, $J_z\!=\!J_y\!=\!0$, and $B\!=\!0$, within the chains. (e,f) Correlation analysis for layer 1 and layer 2, respectively, for the exchange interaction. (g) Correlation analysis for layer 1 for the Ising interaction case (layer 2 all correlations are identically zero). The lower red curves corresponds to the tripartite global quantum discord, $D_3$, and the upper blue curves is the mutual information, $I_3$ in each panel. (h) Rescaled mutual information, $\overline{\mathcal{I}}$, between the system and the environment when the complete chains are taken as indivisible fragments, regardless of interaction type.}
\label{fig2_trot}
\end{figure*}
\subsection{Total Mutual Information and Global Quantum Discord}
In order for objectivity to emerge, strong correlations must be established throughout the compound. However, the nature of these correlations, i.e. whether they are quantum or classical, is important~\cite{ZurekSciRep2013, GiorgiPRA2015}. We will be interested in studying how quantum and classical correlations are established within a given layer in order to provide insight into precisely why redundancy can emerge on different timescales within the layers. For a given $n$-partite state $\rho_\text{tot}$ we will consider the total mutual information~\cite{kavanPRL2010}
\begin{equation}
I_n=\sum_{l=1}^n S(\varrho_{l})-S(\rho_\text{tot}),
\end{equation}
where $\varrho_{l}$ corresponds to the reduced density matrix for the $l^{\rm th}$ subsystem. This quantity encapsulates all correlations, both classical and quantum, present in the state. We will also consider the global quantum discord~\cite{RulliPRA}, a quantity that captures purely quantum correlations defined as
\begin{equation}
    D_n=\underset{\Pi}{\text{min}}[S(\rho_\text{tot}||\Pi_\text{tot} \rho_\text{tot} \Pi_\text{tot}) -\sum_{l=1}^{n}S(\varrho_{l}||\Pi_{l} \rho_\text{l} \Pi_l )].
\end{equation}
Here, $ S(\sigma \| \chi)\!=\!\text{Tr}\left[\sigma \ln \sigma - \sigma \ln \chi \right]$ is the quantum relative entropy and $\Pi_l\!=\!\left\{ \ketbra{l}{l} \right\}$ are the associated projection operators~\cite{RulliPRA, CampbellNJP, CampbellQIP}. It should be noted that for the case of bipartite systems, these quantities become equivalent to the bipartite mutual information and quantum discord, respectively. 

The complementarity between classical and quantum correlations was extensively explored in Ref.~\cite{ZurekSciRep2013, Le2018}, where it was shown that, in order for objectivity to emerge, the genuine quantum correlations between system and environmental fragment should vanish, while the accessible classical information should be as large as possible. Here we follow a similar line, however examining how the correlation profile within the individual layers of the onion model affect the emergence of objectivity.

\section{Quantum Darwinism in the Onion Model}
\label{DarwinInTheOnion}
We begin considering a situation wherein the system and environments are initialised in the state 
\begin{equation}
\label{initialstate}
\ket{\psi}_{S\mathcal{E}}\!=\!\ket{1}_S \bigotimes \ket{0_1\dots0_L}^{\otimes N}
\end{equation}
and the system-environment interaction Eq.~\eqref{SysEnvInt} takes place for a time up to $t\!=\!\pi/{4J}$, which corresponds to when perfect  encoding of the system's information is redundantly imprinted on the first layer~\cite{CampbellPRA2019}. If the system-environment interaction is switched off and the sub-environment interactions in Eq.~\eqref{subenvint} are subsequently switched on, taking the full environment chains $E_j$ as fragments implies that redundancy is guaranteed, and the details of the interactions within the chains do not play a role. Indeed, in Fig.~\ref{fig2_trot}(h) we explicitly compute $\overline{\mathcal{I}}$ for both the exchange and Ising interactions (i.e. for $J_x\!=\!J$, $J_z\!=\!0$, $B\!=\!0$ with $J_y\!=\!J$ and 0, respectively), finding an identical behavior when the complete chains are taken as indivisible fragments. However, if we turn our attention to the behavior of individual layers, we find that the specific form that Eq.~\eqref{subenvint} takes determines whether signatures of quantum Darwinism are able to proliferate through the layers. We demonstrate this in Fig.~\ref{fig2_trot} for the minimal setup consisting of three sub-environments, each with two qubits, and we consider two types of interaction within the chains. The red lines present in the contour plots indicate $\overline{\mathcal{I}}\!=\!1$. A plateau at this value for $f\!\in\!(1/3,2/3)$ indicates a redundant encoding of the system information within the considered environmental fragments.

For the exchange interaction, we find that the redundancy plateau that is initially present in the inner layer, Fig.~\ref{fig2_trot}(a) is smoothly exchanged to the outer layer, Fig.~\ref{fig2_trot}(b), and the dynamics persists as such. Conversely, an Ising interaction does not allow for the information to propagate to the outer layer [cf. Fig.~\ref{fig2_trot}(c) and (d)]. These results demonstrate the important role that the nature of intra-environment interactions can have in regards to spreading redundancy through the layers. Specifically, interactions that preserve the total number of excitations allow for the information to move through the sub-environments, while conversely we see interactions such as the Ising enforce a localization of the information within the first layer and it remains essentially trapped. For the exchange interaction case, we analyze the correlations shared among the constituents for the inner and outer layer in Fig.~\ref{fig2_trot}(e) and (f), respectively where the complementarity between the total and quantum correlations is evident. We see that a redundancy plateau emerges for a given layer when the mutual information is maximized and the discord vanishes, indicating that the fragments of a given layer only share strong classical correlations when objectivity is present~\cite{Le2018}. For the Ising interaction, only the inner layer establishes non-zero correlations during the entire dynamics, shown in Fig.~\ref{fig2_trot}(g). While in this case the discord is zero throughout, we again see that the redundancy plateau corresponds to when the total correlations are maximized.

As the net effect of the exchange interaction is to swap the state between the inner and outer layers and vice-versa, we see a transfer of the redundantly encoded information between them. However, it is worth noting that at the level of individual layers, objectivity is only ever observed confined in a single layer at a time. In contrast, while redundancy clearly emerges as expected in the first layer, the Ising-type interaction is unable to proliferate information. While somewhat artificial, this setting demonstrates the delicate nature of revealing objectivity when the environment itself is a complex entity. In what follows, we will show that the qualitative picture established here persists. In particular, we will explicitly demonstrate that quantum Darwinism is exhibited at some point in the dynamics if the complete sub-environments, $E_j$, are themselves taken as indivisible fragments. Furthermore, we will establish that the nature of the interaction within the chains dramatically affects whether redundant encoding emerges more than ephemerally. We remark that although we have considered a minimal model, we expect the qualitative features persist for larger $N$, providing explicit evidence in the following subsection.

\subsection{Exchange Interactions}
\label{ExchangeSec}

\begin{figure*}[t]
\includegraphics[width=2.0\columnwidth]{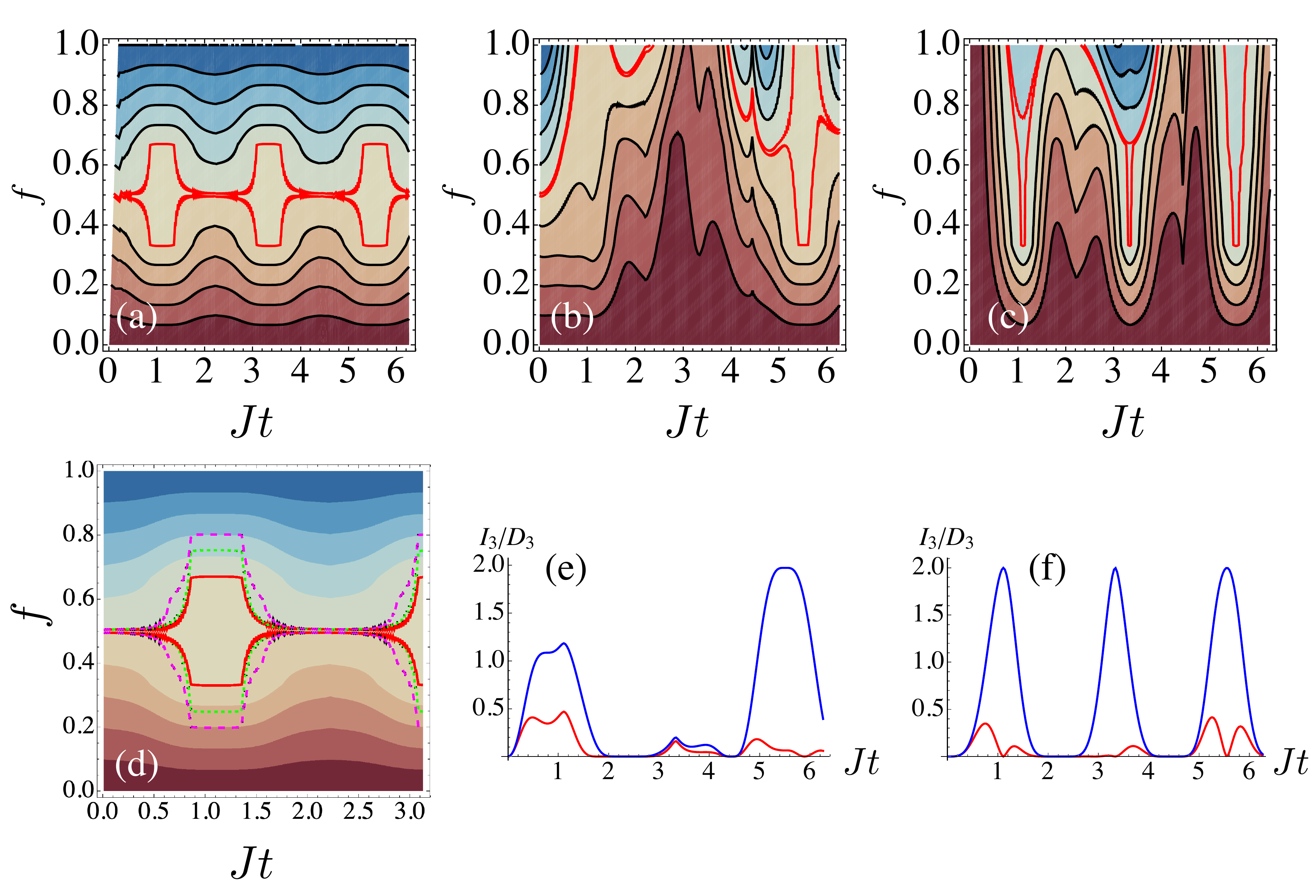}
\caption{Dynamics of $\overline{\mathcal{I}}$ when both interaction terms, Eq.~\eqref{SysEnvInt} and \eqref{subenvint}, are switched on concurrently at $t\!=\!0$ with $J_x\!=\!J_y\!=\!J$, $J_z\!=\!0$ and $B\!=\!0$. (a) Corresponds to the shared information when the complete chains are taken as indivisible fragments. (b,c) Show the rescaled mutual information when the qubits in layer 1 and layer 2 are only considered respectively. In all panels The thick red contour corresponds to $\overline{\mathcal{I}}\!=\!1$ and the contours correspond to the same scaling as shown in Fig~\ref{fig2_trot}. (d) Shows that for an environment consisting of more chains with $N=3$(red), 4 (green dash-dotted) and 5 (magenta dashed), the qualitative features persist and the regions of redundancy start to spread out. (e) Inner layer and (f) outer layer correlation analysis showing tripartite mutual information (upper,blue) and global quantum discord (lower,red).}
\label{normal_exchange}
\end{figure*}
Having established a qualitative understanding of how the nature of the interaction can affect the emergence of Darwinistic features we now consider the more realistic scenario when all interactions occur concurrently. Thus, we begin with the same initial state, Eq.~\eqref{initialstate}, and fix $J_x\!=\!J_y\!=\!J$, $J_z\!=\!0$ and $B\!=\!0$. At $t\!=\!0$ both interaction terms, Eq.~\eqref{SysEnvInt} and \eqref{subenvint} are switched on and the whole system evolves. In Fig.~\ref{normal_exchange}(a) we examine the behavior of Eq.~\eqref{MIeq} when each spin-chain is treated as a fragment. Here we see a redundancy plateau emerges periodically indicating that while qualitatively affecting the the emergence of classical objectivity, provided the ``complete" chain is taken as a fragment quantum Darwinism is still present. In essence, the system-environment interaction is sufficient to establish the correlations necessary for redundant encoding. Fig.~\ref{normal_exchange}(d) we demonstrate that the qualitative features persist for larger environments with evidence of an extensive behavior emerging. Therefore, in what follows we focus on the minimal setup of three chains which is sufficient to capture all the salient features of the model.

Turning our attention to the individual layers, clean signatures of objectivity are largely lost. For the inner layer, in Fig.~\ref{normal_exchange}(b) we find only a single clear instance of a redundancy plateau emerging in the considered time-window near $Jt\!\sim\!5.5$. Curiously however, the second layer, Fig.~\ref{normal_exchange}(c), exhibits more instances of redundant encoding, indicating that the correlations established by the system with the first layer are quickly shared with the outer one. Finally, it is worth noting that in contrast to the staggered interaction case previously considered, here we find instants of time where a redundancy plateau emerges in both layers simultaneously (for instance at $Jt\!\sim\!5.5$). We can understand this by examining the behavior of the total correlations and global quantum discord within each layer as shown in Fig.~\ref{normal_exchange}(e) and (f). We again find that the emergence of redundancy plateaux coincide with when the total correlations are maximized and the discord vanishes.

\subsection{Ising Interactions}
\label{IsingSec}
\begin{figure}[t]
\includegraphics[width=1.05\columnwidth]{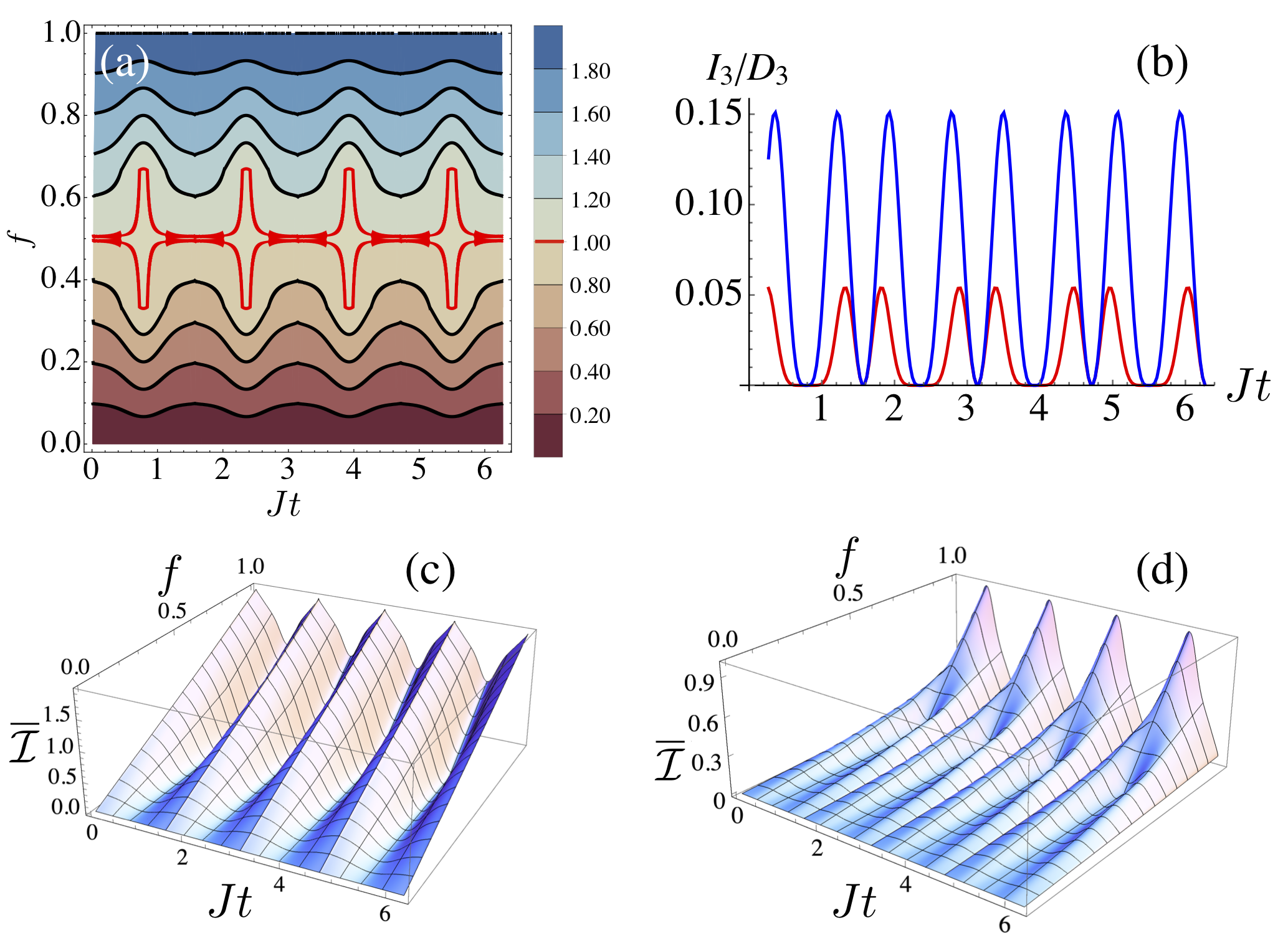}
\caption{Dynamics of ${\overline{\cal I}}$ for an Ising interaction within the chains. (a) The rescaled mutual information between the system and complete chains. (b) Shows the correlations within layer 1. (c,d) Show the rescaled mutual information shared between the system and only the qubits in layer 1 [panel (c)] and layer 2 [panel (d)].}
\label{normal_ising}
\end{figure}

Turning our attention to an Ising-interaction within the chain we fix $J_x\!=\!J$ and $J_z=\!J_y\!\!=\!0$ and $B\!=\!0$ and consider the same initial state. If we again assume all interactions are switched on at $t\!=\!0$ we find that the emergence of a redundancy plateau is periodic when the complete chains are considered as fragments, cfr. Fig.~\ref{normal_ising}(a). However, we now see that the nature of the internal interactions is to suppress these plateaux to sharp temporal windows, which is in marked contrast to the case of the exchange interactions where a redundant encoding persisted for much larger dynamical periods. Fig.~\ref{normal_ising}(c) and (d) we examine the mutual information, $\overline{\mathcal{I}}$, shared between the system and the the inner and outer layer, respectively. It is apparent that redundant encoding is never established for this type of interaction in either layer. Examining the correlation profile within a given layer sheds light onto why neither layer exhibits any Darwinistic features. In Fig.~\ref{normal_ising}(b) we find that the correlations shared among the various constituents of the inner layer remain small and oscillate significantly more frequently, while we have verified that all the qubits in the second layer remain fully uncorrelated and thus never acquire any information about the system.

The previous two case studies clearly demonstrate that the nature of the interactions within a structured environment has a significant effect on the emergence of Darwinistic features. While the Ising interaction continues to allow brief instances of redundant encoding to emerge at the level of complete sub-environments, this is completely lost at the level of individual layers in contrast to the exchange interaction. 

\begin{figure*}[ht]
\includegraphics[width=2\columnwidth]{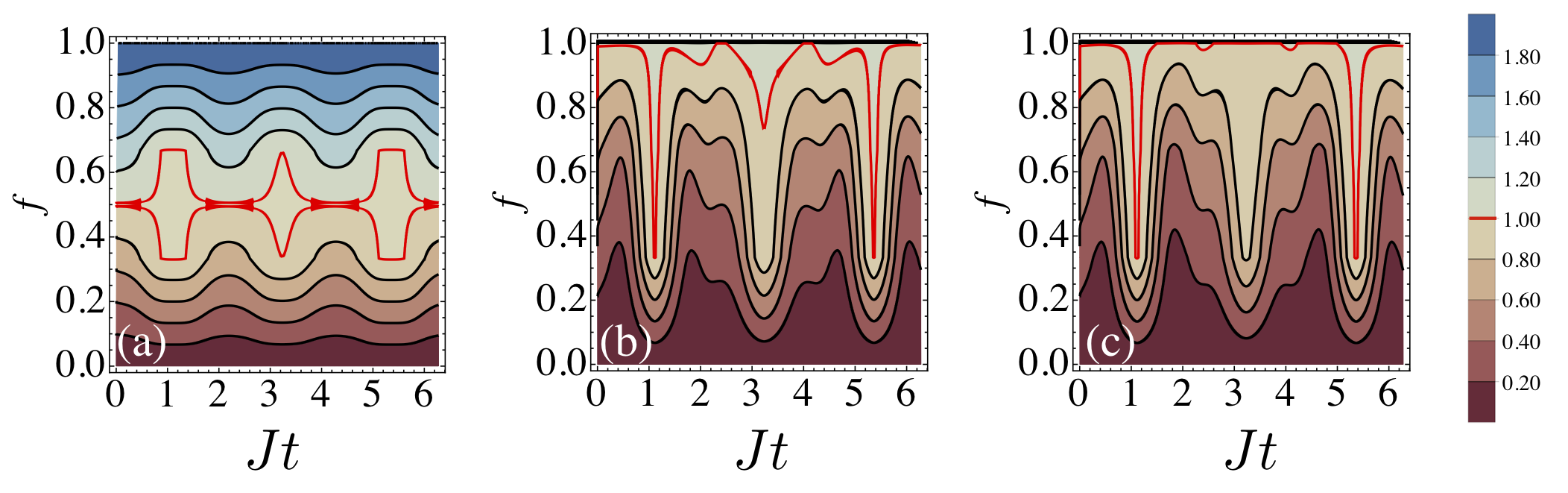}\\
\caption{As shown in panels (a), (b) and (d) of Fig.~\ref{normal_exchange} except the initial state is now $\ket{\psi}_{S\mathcal{E}}\!=\!\ket{1}_S \bigotimes \ket{\psi_{GS}}^{\otimes N}$ where $\ket{\psi_{GS}}$ is the ground state of the chain interaction Hamiltonian. Here we consider an $XX$ type interaction with $J_x\!=\!J_y\!=\!1$, $J_z\!=\!0$ and $B\!=\!0.1$ in Eq.~\eqref{subenvint}.}
\label{GS_exchange}
\end{figure*}
\subsection{Exploring the Role of Different Initial States}
We next consider how the microscopic details of the chains alter the emergence of quantum Darwinism. To this end, we begin with fixing $J_x\!=\!J_y\!=\!J$, $J_z\!=\!0$ in Eq.~\eqref{subenvint}, i.e. we consider an $XX$ interaction within the chains. Furthermore, we assume that the initial state of the chains is the ground state of such an interaction Hamiltonian. For the small chains considered here, there are only two possible ground states, $\ket{\psi_{E_j}}=\ket{01} +\ket{10}/{\sqrt{2}}$, for $0<B/J\!<\!1$ and $\ket{\psi_{E_j}}\!=\!\ket{00}$ for $B\!>\!J$. In Fig.~\ref{GS_exchange} we set $B\!=\!0.1$ we find signatures of quantum Darwinism oscillate. Now we find redundancy plateaux emerge at all levels, i.e. in both layers and complete chains, at the same instants in time. This can be understood due to the strong correlations that are initially present within the chains. At $t\!=\!0$ they form a maximally entangled Bell pair. Switching on the interactions with the system leads to the establishment of correlations between the system and the first layer which necessarily degrades the entanglement within the chains. 

Increasing the strength of the magnetic field, for $B\!>\!J$ the ground state for each chain is $\ket{00}$. When the field is sufficiently strong we find the complete loss of any signatures of quantum Darwinism at all levels (in our simulations $B/J\!=\!2$ was already suitably strong). While the initial states in this case are identical to those discussed in Sec.~\ref{ExchangeSec}, if a strong magnetic field dominates the interaction it suppresses the establishment of any correlations between the system and environment thus leading to a loss of any signatures of quantum Darwinism.

For an Ising interaction, $J_{x}=J,~J_{y}=J_{z}=0$ with a small magnetic field $B/J\! \lesssim \!0.2$, we find a largely consistent behavior as exhibited by the $XX$ interactions at the level of the complete chains. However, unlike in the case of the exchange interaction and despite the ground state being similarly highly entangled for small values of the field, we do not observe any redundancy emerging in the layers, further confirming that that the nature of the environmental interactions greatly effects the emergence of Quantum Darwinism.

\section{Conclusions}
\label{conc}
We have examined whether and how the redundant encoding of information can proliferate within a complex, structured spin environment. Within the framework of quantum Darwinism, where classical objectivity is signalled by a constant mutual information shared between the system and all fragments of the environment, we demonstrated through an extended spin-star model, dubbed ``onion" model, that the microscopic details play an important role. In particular, energy conserving interactions, such as exchange or $XX$, allow for objectivity to emerge and proliferate from the inner-most layer outward. In contrast, interactions such as an Ising interaction between the layers largely suppress the emergence of any Darwinistic features. By examining the total and quantum correlations present within a given layer we were able to establish that the observed behaviors are fully consistent with the presence or absence of quantum correlations, and these results appear to be in line other approaches to exploring the emergence of classical objectivity~\cite{KorbiczPRL2014, HorodeckiPRA2015, KorbiczPRA2017, KorbiczPRA19, LePRA2018, Le2018, Le2020}. Furthermore, while quantum Darwinism provides a framework to explain how classical objectivity arises due to direct system-environment interactions, our work demonstrates that these signatures are not easily spread within the environment itself. In particular, for an energy (or information) preserving interaction there are clear indications that such a proliferation can occur, while other interactions generally suppress the characteristic features of quantum Darwinism, thus suggesting that the proliferation of relevant system information indirectly through an environment may not be a generic feature.

\acknowledgements
E.R. is supported by the Northern Ireland Department for Economy (DfE). M.P. acknowledges support from the H2020-FETOPEN-2018-2020 project TEQ (grant No. 766900), the DfE-Science Foundation Ireland (SFI) Investigator Programme (grant 15/IA/2864), COST Action CA15220, the Royal Society Wolfson Research Fellowship (RSWF$\backslash$ R3$\backslash$183013), the Royal Society International Exchanges Programme (IEC$\backslash$R2$\backslash$ 192220), the Leverhulme Trust Research Project Grant (grant No. RGP-2018-266), and the UK-EPSRC. S.C. acknowledges the SFI Starting Investigator Research Grant ``SpeedDemon" No. 18/SIRG/5508.

\bibliography{mauros_onion}

\end{document}